\documentclass[a4paper,11pt,notoc]{article}
\usepackage{jheppub} 

\usepackage{ulem}
\usepackage{graphicx}

\newcommand{\mDM}{m_{\rm{DM}}}
\newcommand{\mMed}{M_{\rm{med}}}
\newcommand{\gDM}{g_{\rm{DM}}}
\newcommand{\gq}{g_q}

\title{Interplay and Characterization of Dark Matter Searches at Colliders and in Direct Detection Experiments}

\vspace{0.5em}

\author[a]{Sarah~A.~Malik,}
\author[b,c]{Christopher~McCabe,}
\author [a]{Henrique~Araujo,}
\author[d,e]{Alexander~Belyaev,}
\author[b]{C\'eline~B\oe hm,}
\author[f]{Jim~Brooke,}
\author[a]{Oliver~Buchmueller,}
\author[a]{Gavin~Davies,}
\author[g,h]{Albert~De~Roeck,}
\author[a]{Kees~de~Vries,}
\author[i]{Matthew~J.~Dolan,}
\author[g,j]{John~Ellis,}
\author[j]{Malcolm~Fairbairn,}
\author[f]{Henning~Flaecher,}
\author[k]{Loukas Gouskos,}
\author[b]{Valentin~V.~Khoze,}
\author[l]{Greg~Landsberg,}
\author[f]{Dave~Newbold,}
\author[m]{Michele~Papucci,}
\author[a]{Timothy~Sumner,}
\author[d,e]{Marc~Thomas}
\author[e]{and~Steven~Worm}

\affiliation[a]{High Energy Physics Group, Blackett Laboratory, Imperial College, Prince Consort Road, London, SW7 2AZ, UK\ }
\affiliation[b]{Institute for Particle Physics Phenomenology, Durham University, Durham, DH1 3LE, UK\ }
\affiliation[c]{GRAPPA, University of Amsterdam, Science Park 904, 1098 XH Amsterdam, Netherlands\ }
\affiliation[d]{School of Physics and Astronomy, University of Southampton, Highfield, Southampton, SO17 1BJ, UK\ }
\affiliation[e]{Particle Physics Department, Rutherford Appleton Laboratory, Chilton, Didcot, OX11 0QX, UK\ }  
\affiliation[f]{HH Wills Physics Laboratory, Tyndall Avenue, Bristol, BS8 1TL, UK\ }
\affiliation[g]{Physics Department, CERN, CHÐ1211 Gen\`eve 23, Switzerland\ }
\affiliation[h]{Antwerp University, BÐ2610 Wilrijk, Belgium\ }
\affiliation[i]{Theory Group, SLAC National Accelerator Laboratory, Menlo Park, CA 94025, USA\ }
\affiliation[j]{Theoretical Particle Physics and Cosmology Group, Department of Physics, King's College London, London, WC2R 2LS, UK\ }
\affiliation[k]{University of California, Santa Barbara, Department of Physics Broida Hall, Bldg.~572, Santa Barbara, CA 93106-9530, USA\ }
\affiliation[l]{Physics Department, Brown University, Providence, Rhode Island 02912, USA\ }
\affiliation[m]{Lawrence Berkeley National Laboratory, Berkeley, CA 94720, USA\ }

\abstract{In this White Paper we present and discuss a concrete proposal for the consistent interpretation of Dark Matter searches at colliders and in direct detection experiments. Based on a specific implementation of simplified models of vector and axial-vector mediator exchanges, this proposal demonstrates how the two search strategies can be compared on an equal footing. \\ \\
{\it White Paper from the Brainstorming Workshop held at Imperial College London on May 29th, 2014. A link to the Workshop's agenda is given in~\cite{Agenda}.}\\ \\
{\tt IPPP/14/83, DCPT/14/166, KCL-PH-TH/2014-37, LCTS/2014-36, CERN-PH-TH/2014-180}
}

\begin{document}
\maketitle
\flushbottom

\section{Scope of the Workshop}
\label{sec:scope}

Since the start-up of the LHC in 2010, collider searches for Dark Matter (DM) particle
production, and their comparison with direct detection (DD) scattering experiments such as XENON100~\cite{Aprile:2011dd} and LUX~\cite{Akerib:2012ys}, 
have become a focal point for both the experimental and theoretical particle and astroparticle communities.    

Collider searches are generally characterized by their use of `mono-objects', such as mono-jets or mono-photons,
accompanied by missing transverse energy~\cite{Boehm:2003hm,Cao:2009uw,Beltran:2010ww,Goodman:2010yf,Bai:2010hh,Goodman:2010ku,Rajaraman:2011wf,Fox:2011pm}.
Until recently, these searches were mainly interpreted in the framework of specific models,
such as the ADD~\cite{ArkaniHamed:1998rs} or unparticle models~\cite{Georgi:2007ek},
or else used an effective field theory (EFT) to allow for the straightforward comparison with the results of
DD experiments.

However, interpretations within specific models are often too narrow in scope,
and several independent 
groups~\cite{Bai:2010hh,Fox:2011pm,Goodman:2011jq,Shoemaker:2011vi,Buchmueller:2013dya,Busoni:2013lha,Busoni:2014sya,Busoni:2014haa}
have pointed out that the interpretation within the EFT framework can lead to the wrong conclusions when comparing collider results with the results from DD experiments.  As an alternative, a simplified model
description of collider and DD searches has been advocated in order to avoid these pitfalls~\cite{An:2012va,Frandsen:2012rk,1303.3348, 1307.8120, 1308.0592, 1308.0612, 1308.2679,Buchmueller:2013dya, 1402.2285}.

The Brainstorming Workshop contributed to the development of a consistent simplified framework to interpret these searches,
so as to facilitate comparison of the sensitivities of collider and DD experiments. This is required in order to establish quantitatively the complementarity of these two search approaches,
which is critical in our continuing quest for DM.  A link to the Brainstorming Workshop's agenda, which includes links to the individual talks, is given in~\cite{Agenda}. 
 
In this White Paper, we propose benchmark scenarios in a particular simplified model framework for DM models
and provide examples of plots that will allow for a more meaningful comparison of the results from collider and DD experiments. These
scenarios are summarized in Section~\ref{sec:summary}. This proposal should be considered as a first practical step in the
discussion towards a more complete analysis strategy to be developed in the future.

\section{Comparison of DM searches}
\label{sec:Intro}

Although the Workshop touched on several interesting aspects related to models of DM
and the characterization of DM searches, its main focus was on defining a concrete proposal for
how to go beyond the problematic comparison of DM searches in the EFT framework. Therefore
in this document, we focus mainly on the outline of our proposal for comparing collider and DD searches for
DM on an equal footing, so as to better understand and exploit their complementarity. 
This is largely based on the results of a recent paper~\cite{Buchmueller:2014yoa}
by several of the Workshop participants, whose work was in part inspired by the Workshop.   

While the EFT framework is a convenient tool for interpreting DM searches from DD experiments,
recent work by several independent groups~\cite{Bai:2010hh,Fox:2011pm,Goodman:2011jq,Shoemaker:2011vi,Buchmueller:2013dya,Busoni:2013lha,Busoni:2014sya,Busoni:2014haa} has highlighted the problem
that the EFT interpretation of collider searches suffers from several significant limitations,
which prevent a comprehensive characterization of these searches. 
A comparison of DM searches at collider and DD experiments using the EFT approach 
does not provide an accurate description of the complementarity of the two search strategies.

\subsection{Simplified DM models}

An alternative to the EFT interpretation is the characterization of DM searches using simplified 
models~\cite{Alwall:2008ag,Alves:2011wf}. Simplified models are widely used to interpret missing-energy searches
at colliders in the context of supersymmetry, and have become a successful way to benchmark and
compare the reaches of these collider searches. In contrast to the EFT ansatz,
simplified models are able to capture properly the relevant kinematic properties of collider searches with only a few free parameters. 

As pointed out in~\cite{An:2012va,Frandsen:2012rk,1303.3348, 1307.8120, 1308.0592, 1308.0612, 1308.2679,Buchmueller:2013dya, 1402.2285, Buchmueller:2014yoa}, simplified models of DM also provide an appropriate framework for comparing and characterizing the results of DM searches at colliders and DD experiments.
This was demonstrated within a framework of Minimal Simplified Dark Matter (MSDM) models
with vector and axial-vector mediators exchanged in the s-channel~\cite{Buchmueller:2014yoa}.
While the collider phenomenology of the vector and axial-vector mediators is similar, at DD experiments they are very different. These two cases therefore demonstrate how to compare DD and collider results on an equal footing for two distinctive scenarios. Although these two mediator cases already cover a significant variety of interesting DM models, as we discuss below in more detail, 
it will be important to also consider t-channel exchanges as well as scalar and pseudo-scalar mediators in the future.

The MSDM models are constructed using four parameters: the mass of the DM particle, $\mDM$,
the mass of the mediator, $\mMed$, the coupling of the mediator to the DM particles, $\gDM$, 
and the coupling of the mediator to quarks, $\gq$. For the latter, as a simplifying assumption, the mediator is assumed to couple to all quark flavours with equal strength. In this White Paper we assume that the DM particle is a Dirac fermion~($\chi$) and the new Lagrangian terms for the vector ($Z'$) and axial-vector ($Z''$) MSDM models are
\begin{align*}
\label{eq:AV} 
\mathcal{L}_{\mathrm{vector}}&\supset \frac{1}{2}\mMed^2 Z'_{\mu} Z'^{\mu} - \gDM Z'_{\mu} \bar{\chi}\gamma^{\mu}\chi -\sum_q \gq Z'_{\mu} \bar{q}\gamma^{\mu}q \\
\mathcal{L}_{\rm{axial}}&\supset \frac{1}{2}\mMed^2 Z''_{\mu} Z''^{\mu} - \gDM Z''_{\mu} \bar{\chi}\gamma^{\mu}\gamma^5\chi -\sum_q \gq Z''_{\mu} \bar{q}\gamma^{\mu}\gamma^5q
\end{align*}
where the sum extends over all quarks.

 It is important to emphasize that these four variables represent the minimum set of parameters necessary for the comparison of collider and DD experiments. Direct detection experiments are sensitive
 only to a specific combination of these parameters that enter the nucleon-DM scattering cross section, namely
 $$\sigma^0_{\rm{DD}} \sim \frac{\gDM^2\gq^2\mu^2}{\mMed^4} \, ,$$
 where $\mu$ is the reduced mass of the nucleon-DM system, which asymptotically becomes constant for heavy DM particles. In comparison, all four parameters play different and important roles in collider searches:
 \begin{itemize}
 \item $\mDM$: collider limits depend on~$\mDM$, 
 with the sensitivity limited by the available energy in the centre-of-mass frame;
 \item $\mMed$: the interplay between $\mMed$ and $\mDM$ is very important for sufficiently light mediators,
 as for $\mDM < \mMed/2$ one expects a resonant enhancement of the collider sensitivity to DM;
 \item $\gDM$, $\gq$: the cross section for DM production in collider experiments is sensitive to the
 product of the two couplings squared, as is the DM-nucleon interaction cross section in DD experiments.
 However, in addition, collider experiments are also sensitive to the sum of these couplings squared,
 which determines the width of the mediator~($\Gamma_{\rm{med}}$). If the latter is too large ($\Gamma_{\rm{med}}\gtrsim\mMed$), single-mediator exchange does
 not provide a realistic description of either DM-nucleon scattering or collider production of a pair of DM particles~--- a fact 
 that is often overlooked in the interpretation and comparison of the searches.
\end{itemize} 

To produce the collider limits in MSDM models, we generate events for the DM signal at the LHC using an extension of POWHEG BOX~\cite{Haisch:2013ata,Nason:2004rx,Frixione:2007vw,Alioli:2010xd}. The program generates the process of a pair of DM particles produced in association with a parton at next-to-leading order (NLO). It can be matched consistently to a parton shower, which as discussed in~\cite{Haisch:2013ata}, is of particular importance to simulate accurately the case where jet vetoes are applied in the analysis. This is the case in the monojet analysis where the third jet in the event is vetoed. 
In our case, we match to Pythia 8.180~\cite{Sjostrand:2007gs,Sjostrand:2006za} and put through Delphes~\cite{deFavereau:2013fsa, Ovyn:2009tx} for the detector simulation.

The inclusion of NLO corrections reduces the dependence on the choice of renormalisation and factorisation scales and thereby the theoretical uncertainty, which will become important if a small excess is observed. The program has three further advantages. Firstly, it can generate events for both the EFT case and also simplified models. Secondly, in addition to the vector and axial-vector mediators considered here, it can also be used for studies of scalar mediators. Thirdly, it includes K-factors which are particularly important in models where the scalar couples to gluons in the s-channel.

As demonstrated in~\cite{Haisch:2013fla}, the inclusion of higher order corrections can also be advantageous in probing the structure of couplings between DM and SM, which can be determined by looking at the azimuthal difference between two jets in events where the final state contains two jets together with missing transverse energy.  For instance, for loop-mediated interactions with gluons where a spin-0 particle is exchanged in the s-channel, the CP nature of the latter can be tested.

\subsection{Comparisons of collider and DD limits}

A comprehensive comparison of the limits from collider and DD searches in all of the four 2D projections 
of the 4-parameter MSDM model is provided in~\cite{Buchmueller:2014yoa}. It shows that, 
for the exchange of a vector mediator, only for very light DM masses $(\lesssim5~\mathrm{GeV}$) do
LHC mono-jet searches (represented by the CMS mono-jet search~\cite{CMS-PAS-EXO-12-048, Khachatryan:2014rra})
have  better sensitivity than DD searches (represented by the LUX 2013~\cite{Akerib:2013tjd} and 
SuperCDMS~\cite{Agnese:2014aze} results). For larger DM masses the DD experiments provide significantly 
stronger bounds on the parameter space of vector mediators. For axial-vector mediators, however, the
LHC and DD searches generally probe complementary regions in the full parameter space, 
with the LHC searches having greater sensitivity than the DD experiments for DM masses below around 200~GeV. 

\begin{figure}[t!]
\centering
\includegraphics[width=0.495\columnwidth]{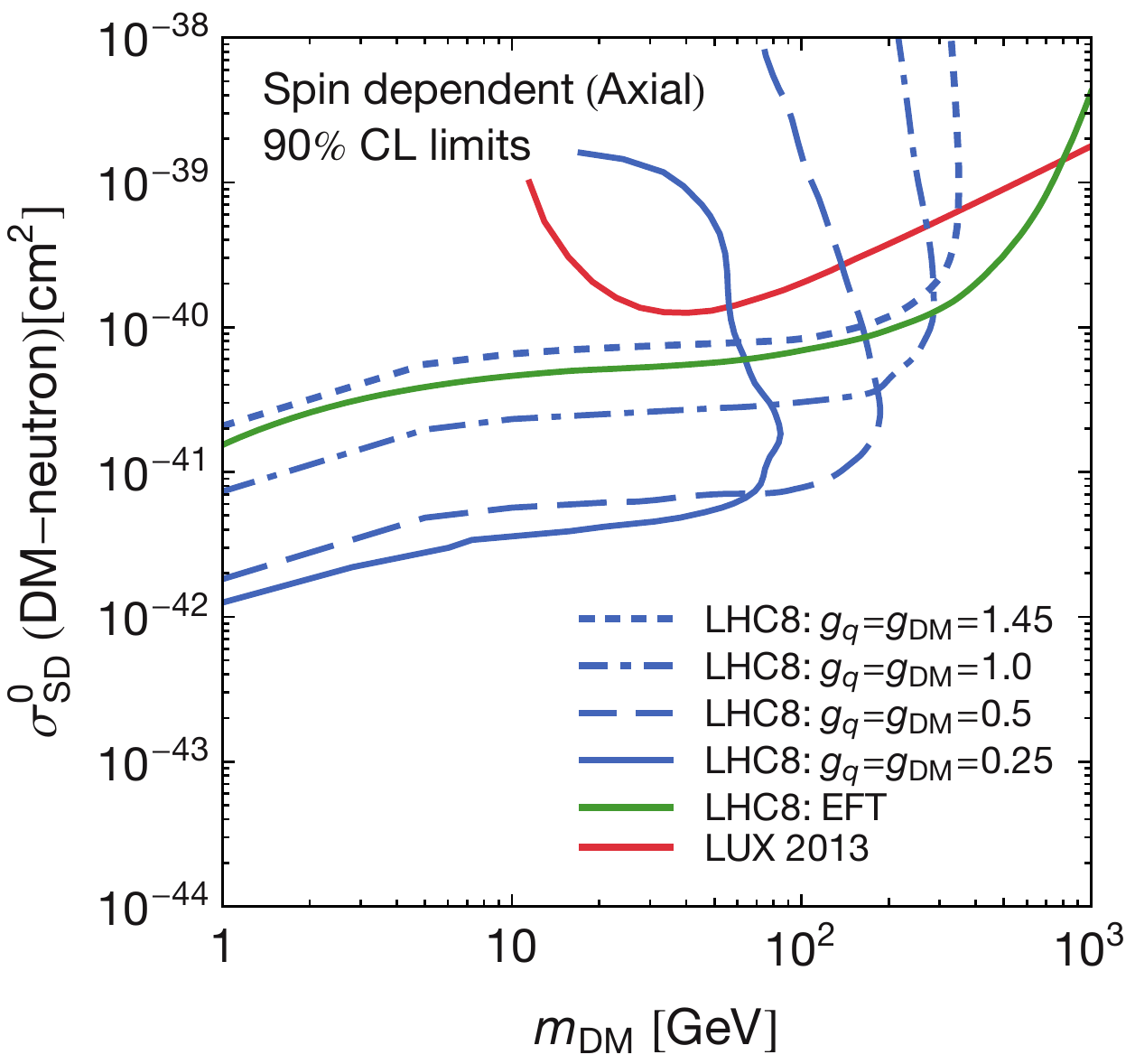} 
\includegraphics[width=0.495\columnwidth]{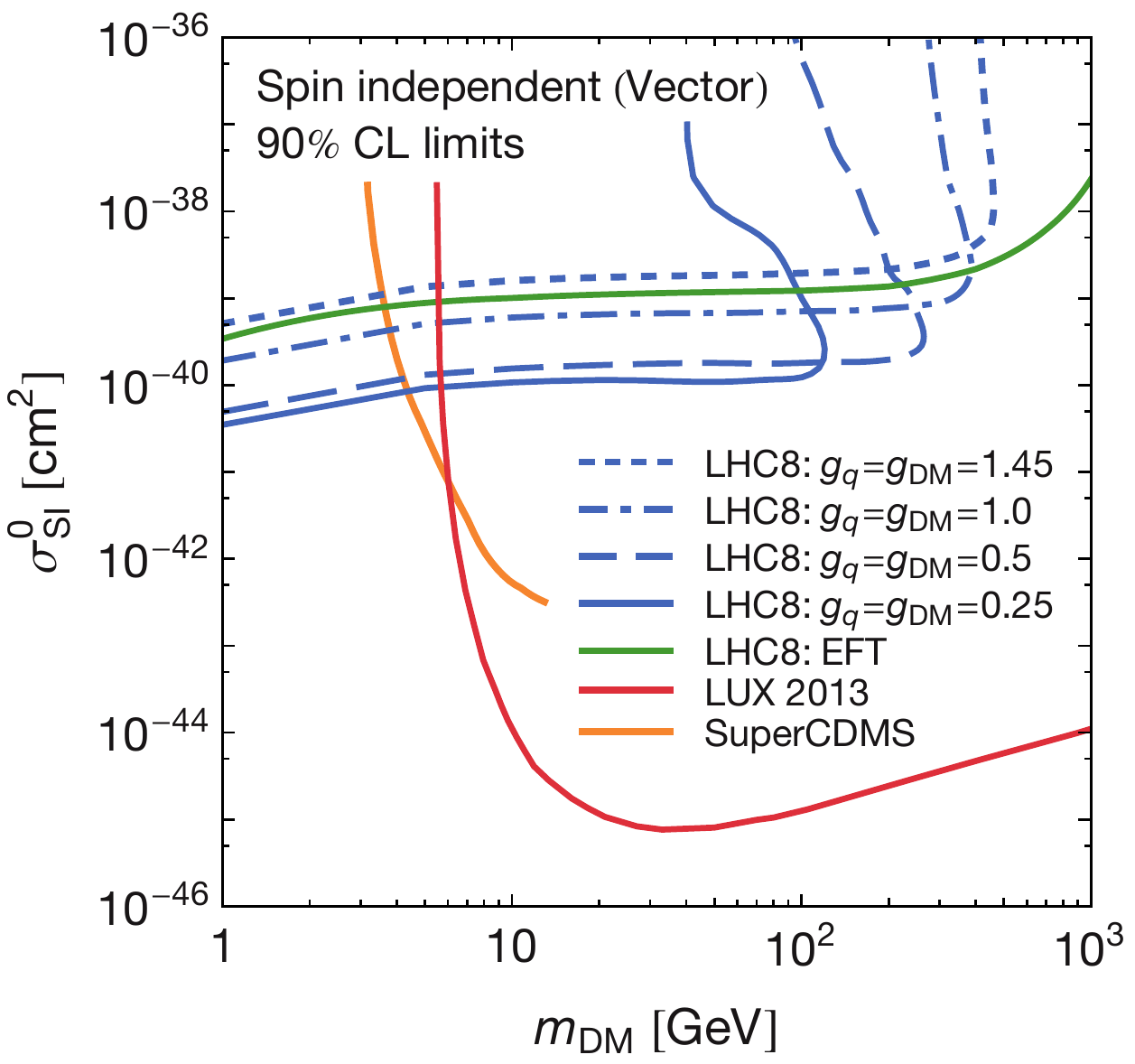}
\caption{A comparison of the current 90\% CL LUX and SuperCDMS limits (red and orange lines, respectively), 
the mono-jet limits in the MSDM models (blue lines) and the limits in the EFT framework (green line) in the 
cross section vs $\mDM$ plane used by the direct detection community. The left and right panels show the limits on the 
SD and SI cross sections appropriate for axial-vector and vector mediators respectively. For the MSDM models we show
scenarios with couplings $\gq =\gDM = 0.25, 0.5, 1.0, 1.45$.}
\label{fig:XS}
\end{figure}   

Traditionally, DD experiments display their results in terms of the DM-nucleon spin-independent and spin-dependent cross sections 
$\sigma^0_{\rm{SI}}$ and $\sigma^0_{\rm{SD}}$, respectively. It is thus also useful to provide
comparisons of the MSDM limits from the mono-jet and DD searches in the $(\sigma^0_{\rm{SI}}, \mDM)$
and $(\sigma^0_{\rm{SD}}, \mDM)$ planes. As discussed in Section~5 of our main reference~\cite{Buchmueller:2014yoa}, 
for fixed couplings $\gq$ and $\gDM$, 
collider limits defined in the $(\mMed, \mDM)$ plane of the MSDM model can be directly translated into the
$(\sigma^0_{\rm{}}, \mDM)$ planes. Vector and axial-vector mediators lead to spin-independent and 
spin-dependent interactions in DD experiments, respectively. For DD searches the cross section scales exactly like
$(\gq\gDM)^2/\mMed^4$, while for collider searches it scales approximately like $(\gq\gDM)^2/(\mMed^4 \Gamma_{\rm{med}})$. 
For small values of the width, as in weakly coupled scenarios, there is a resonant enhancement of the cross section in the collider case.  

Figure~\ref{fig:XS} shows  the MSDM limits from the CMS mono-jet search for different coupling scenarios in the
$(\sigma^0_{\rm{SD}}, \mDM)$ and $(\sigma^0_{\rm{SI}}, \mDM)$ planes (left and right, respectively). 
The MSDM limits for the axial-vector mediator are displayed in the spin-dependent plane, 
and the results from the vector mediator study are shown in the spin-independent plane. 
To assess the dependence of the collider limits on the choice of couplings, four different coupling scenarios are shown:
$\gq = \gDM$ = [0.25, 0.5, 1.0, 1.45] (blue lines).  The two extreme scenarios of 0.25 and 1.45 are  chosen 
because they approximate the range over which the LHC mono-jet search can place meaningful 
limits in the MSDM models. For $\gq = \gDM \gtrsim1.45$ the width of a vector or axial-vector mediator 
exchanged in the s-channel becomes larger than its mass, making a particle physics interpretation of 
the interaction problematic. For $\gq = \gDM \lesssim0.25$ the 8~TeV CMS mono-jet search
no longer has sufficient sensitivity to place a significant limit on the parameter space.   

Figure~\ref{fig:XS} also shows the limit obtained from an interpretation of the mono-jet search in the 
framework of the EFT (green line). The EFT limits should agree with the MSDM limit in the 
domain where the EFT framework is valid. We see that it is only for the extreme coupling scenario
$\gq = \gDM =1.45$ that the EFT limit approximates the MSDM limit, and only for DM masses below around 300~GeV.
For larger $\mDM$ the EFT fails to describe any of the coupling scenarios. 
For weaker couplings, the MSDM limits get stronger for DM masses below around 50 to 300~GeV,
due to the resonant enhancement of the cross section for a s-channel mediator that was explained above.
This effect is absent within the EFT framework. The reach in DM mass of the MSDM limits increases with larger couplings.
Overall, this comparison of the EFT and MSDM limits demonstrates again that the EFT framework is unable to 
capture all of the relevant kinematic properties of the collider searches, which is demonstrated by the large 
disparity between the EFT and MSDM limits. Comparing EFT collider limits with those of DD searches gives a 
misleading representation of the relative sensitivity of the two search strategies, 
especially for weaker coupling scenarios and $\mDM \gtrsim 300$~GeV.  

Finally Figure~\ref{fig:XS} also shows the LUX limits for both interactions (red lines) 
and the spin-independent SuperCDMS limit (orange line). 
Whilst the comparison of the DD search result with the EFT collider limit is biased, 
a comparison with the MSDM limits from the LHC mono-jet analysis, 
which properly describes the kinematic properties of the collider search,
represents a comparison of collider and DD experiments on an equal footing,
establishing quantitatively the complementarity of the two search strategies.
With the exception of light DM masses $\mDM \lesssim 5$~GeV, 
DD experiments provide much stronger limits for vector-mediated interactions.
In the axial-vector case, the collider limits are generally stronger for $\mDM \lesssim 300$~GeV.
This is especially true for small couplings where the collider cross section is further enhanced by the small mediator width.
Owing to the kinematic constraint $\mMed\ge 2\mDM$ on s-channel mediator production at the collider,
DD searches are today the only searches providing significant limits for either cross section for $\mDM\gtrsim$~300 GeV.

\subsection{Future experiments  and upgrades - projected sensitivities}
\label{sec:projection}

With the DD experiments and the LHC programme gearing up for major upgrades,
we also look at their projected sensitivity to DM particles in the future. 
We explore three scenarios for the LHC: 30~fb$^{-1}$ at 13~TeV to gauge the reach of
LHC Run 2, 300~fb$^{-1}$ at 13 TeV to provide an estimate of the reach of
LHC Run 3, and 3000~fb$^{-1}$ at 14 TeV to show the expected reach of the high-luminosity upgrade of the LHC.  
For the DD experiments we show the estimated limit for the lifetime exposure of two liquid xenon experiments: the LUX-ZEPLIN (LZ) experiment~\cite{Malling:2011va},
with an exposure of 10 tonne~years,\footnote{A similar exposure will also be reached by XENONnT~\cite{Aprile:2014zvw}.} and DARWIN~\cite{Baudis:2012bc,Schumann:2015cpa}, with an exposure of 200 tonne~years. We also show the discovery reach for DD experiments when limited by the coherent neutrino scattering background~\cite{Billard:2013qya}. 

\begin{figure}[t!]
\centering
\includegraphics[width=0.495\columnwidth]{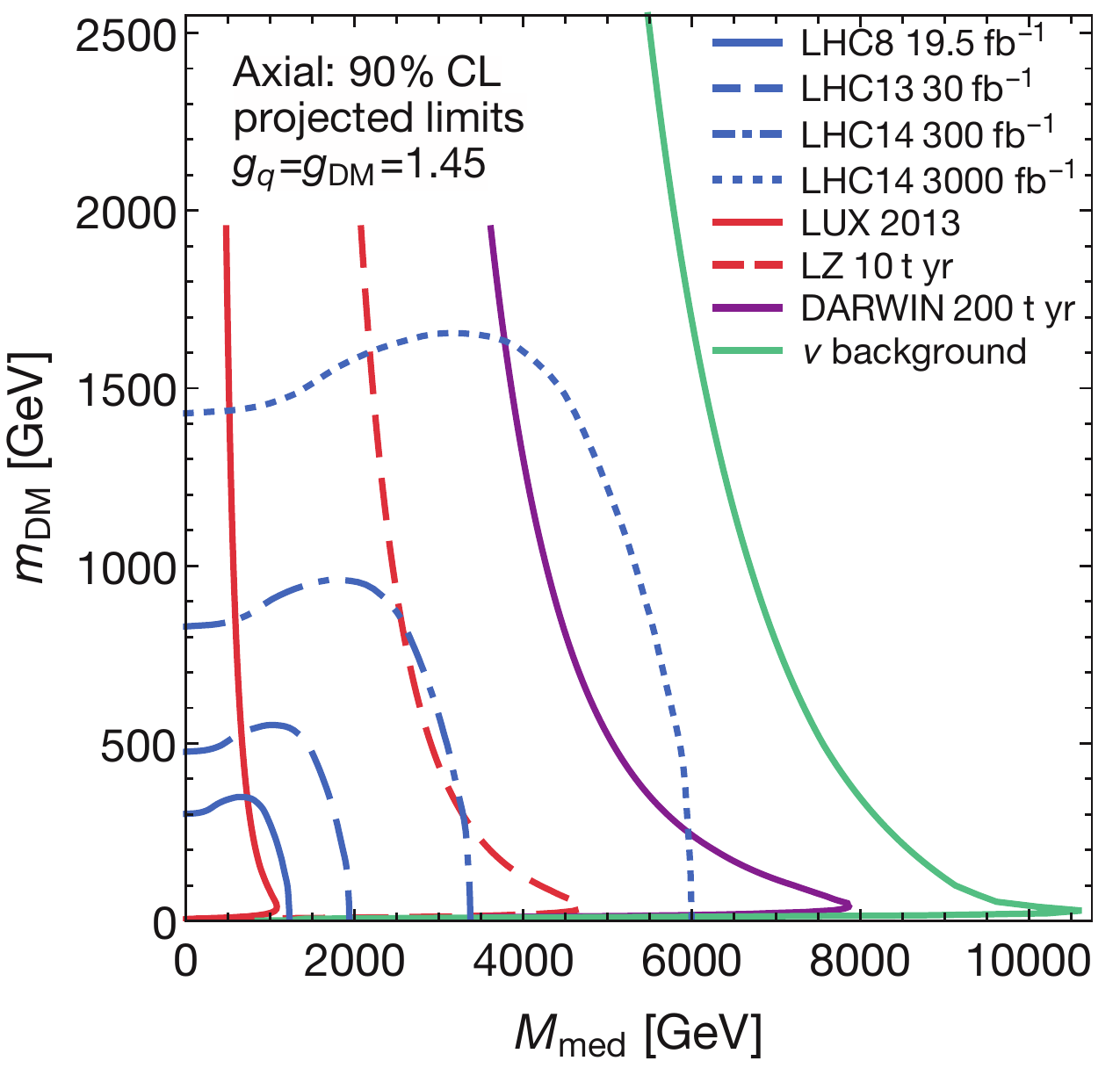} 
\includegraphics[width=0.495\columnwidth]{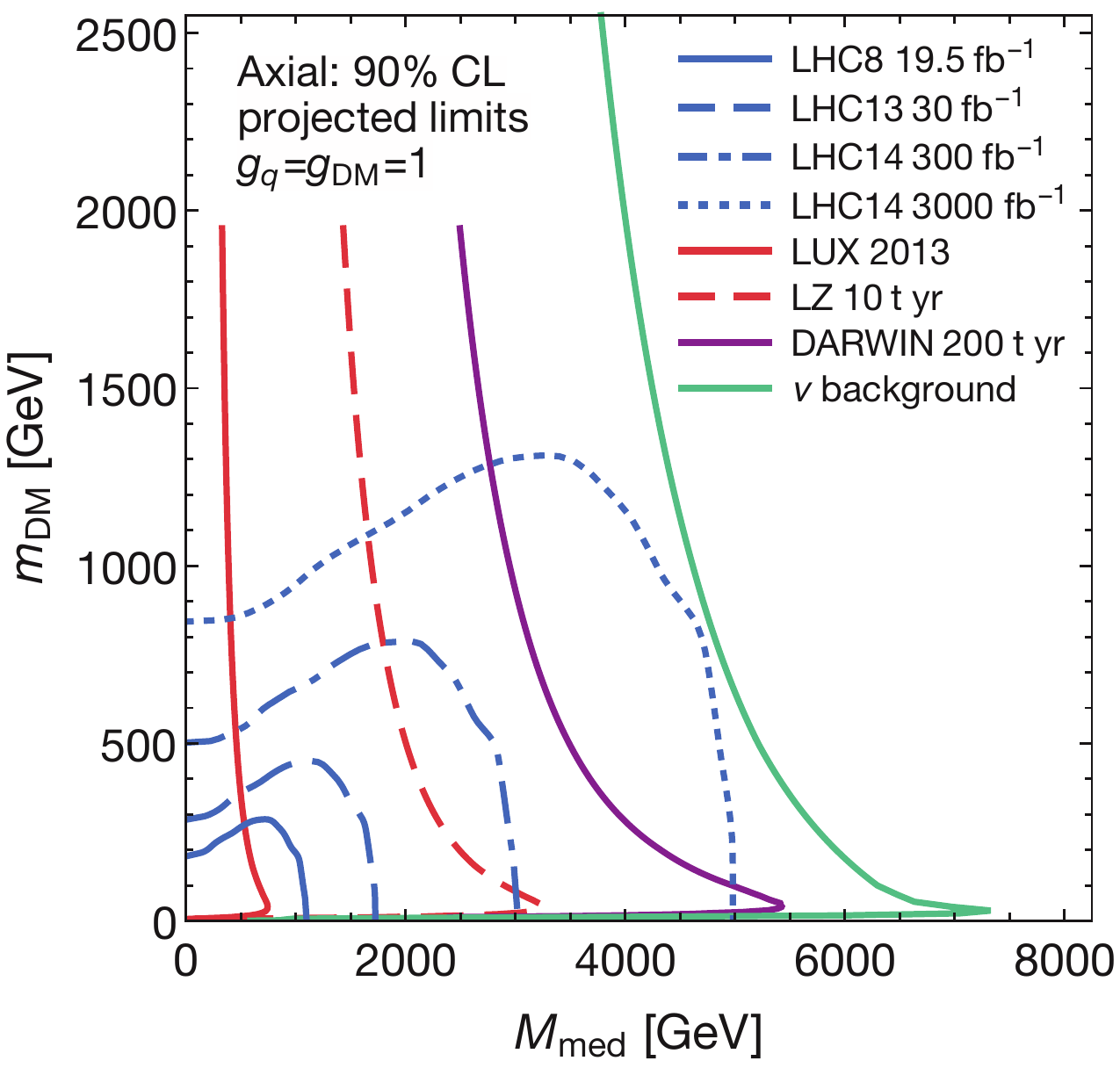}
\includegraphics[width=0.495\columnwidth]{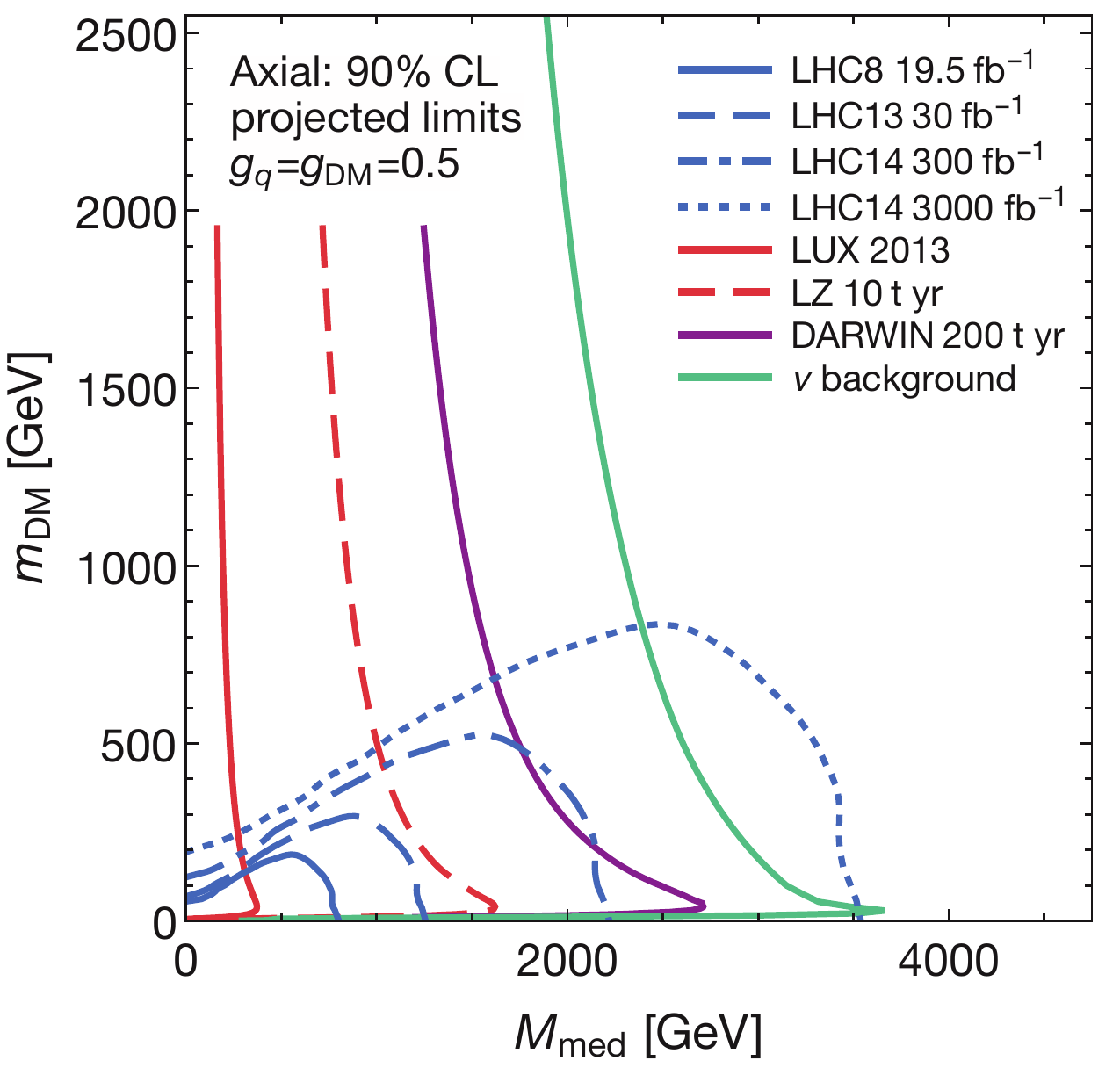}
\includegraphics[width=0.495\columnwidth]{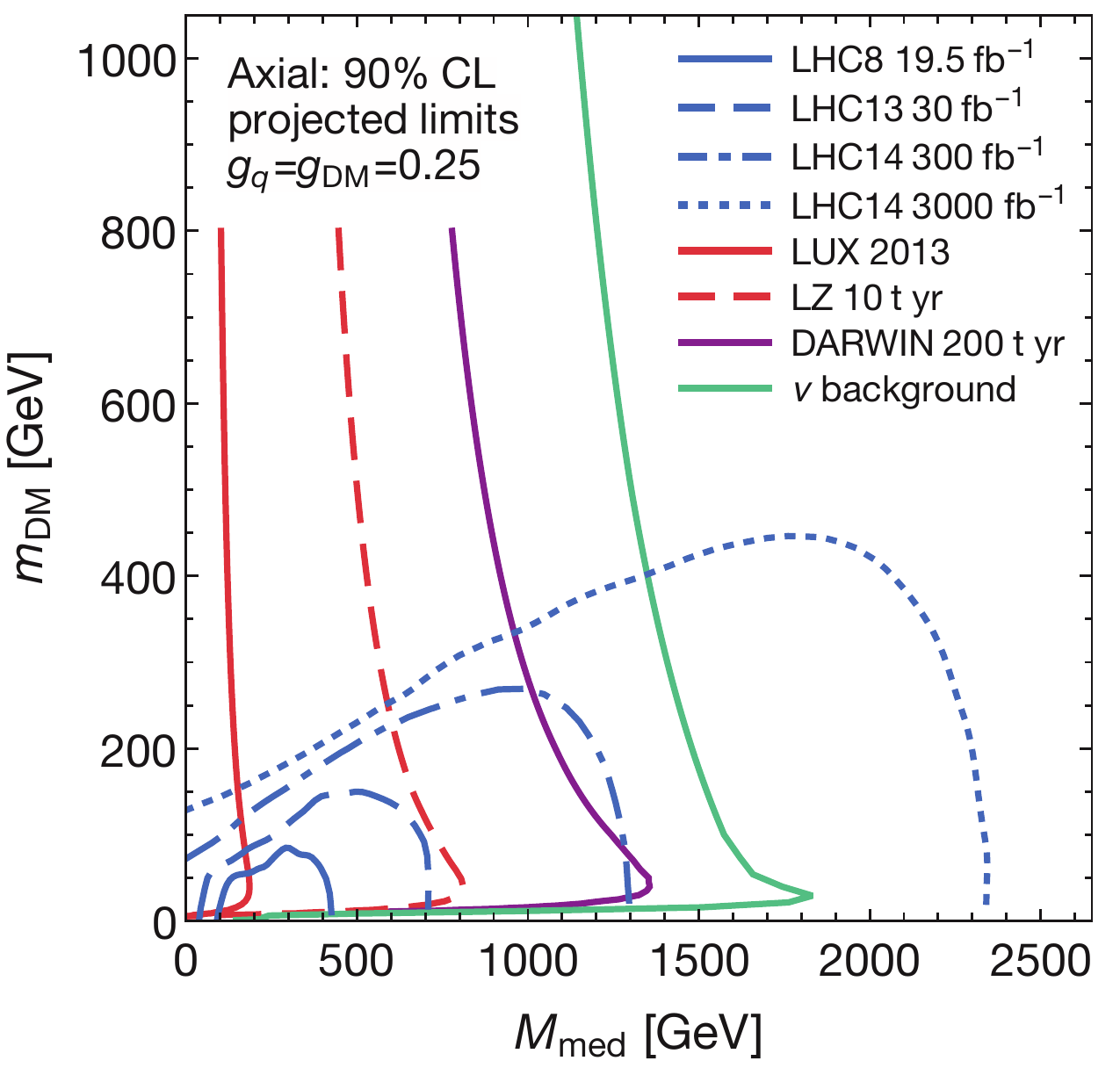}
\caption{Projected limits for the CMS mono-jet search (blue lines) and DD searches by LUX (red line), 
LZ (red dashed line) and DARWIN (purple line) in the
$(\mMed, \mDM)$ plane for an axial-vector mediator with the coupling scenarios
$\gq = \gDM = 0.25$, 0.5, 1.0, 1.45. For reference, the discovery reach of DD experiments accounting for the coherent 
neutrino scattering background is also displayed (green line). The region to the left of the various curves is
excluded at 90\% CL. Note the change in scale in each panel.}
\label{fig:projAV}
\end{figure}

\begin{figure}[t!]
\centering
\includegraphics[width=0.495\columnwidth]{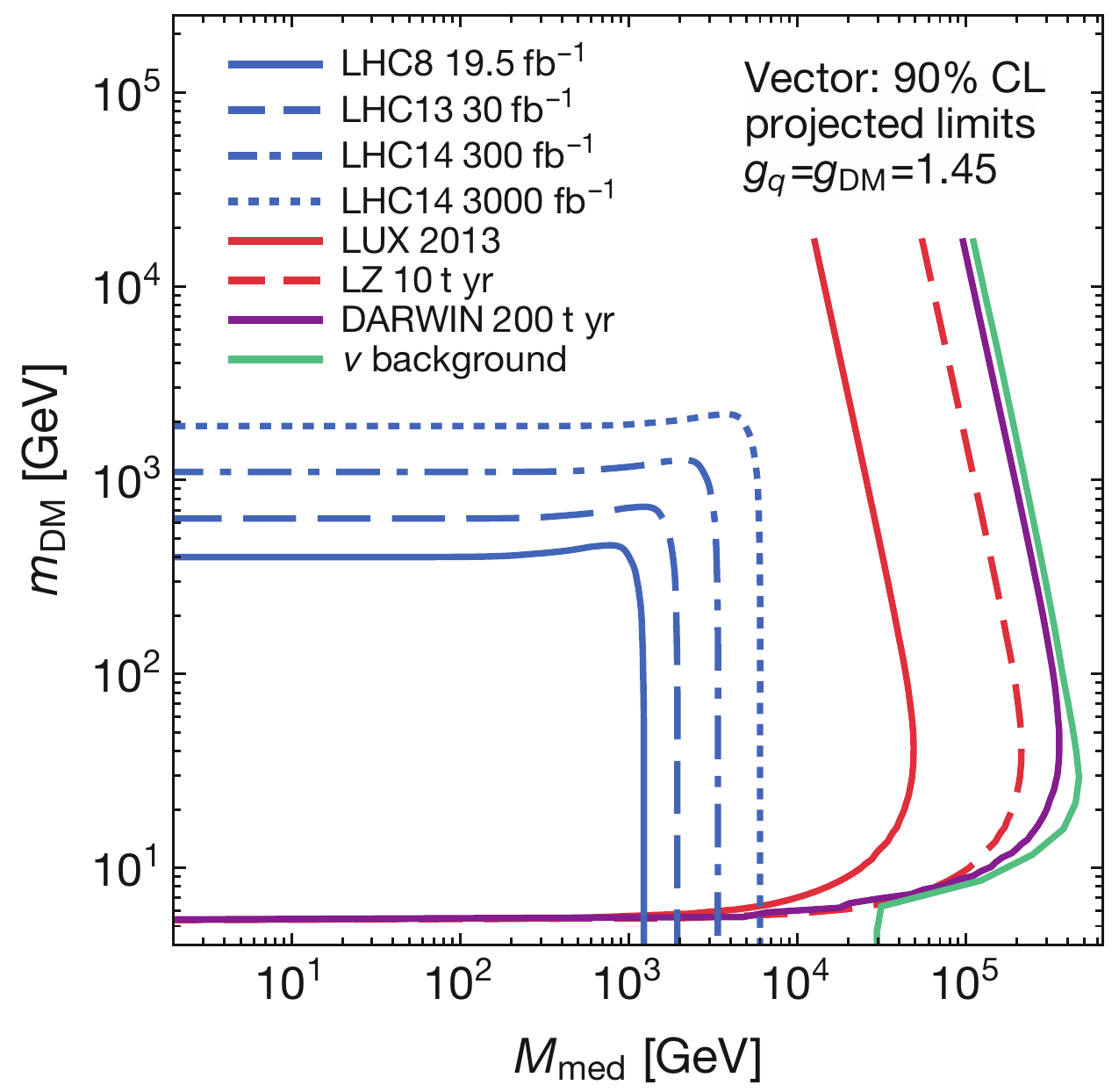} 
\includegraphics[width=0.495\columnwidth]{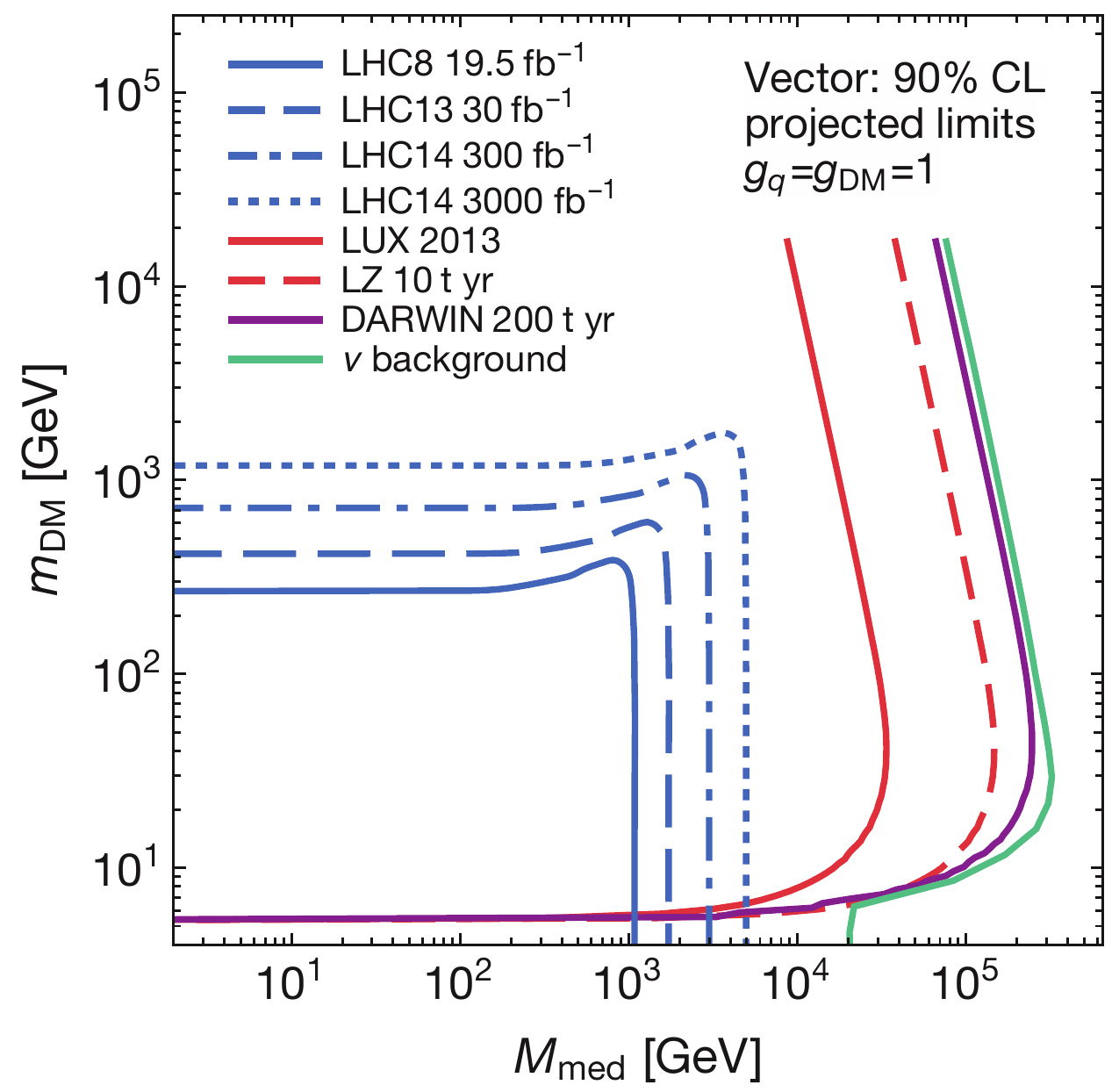}
\includegraphics[width=0.495\columnwidth]{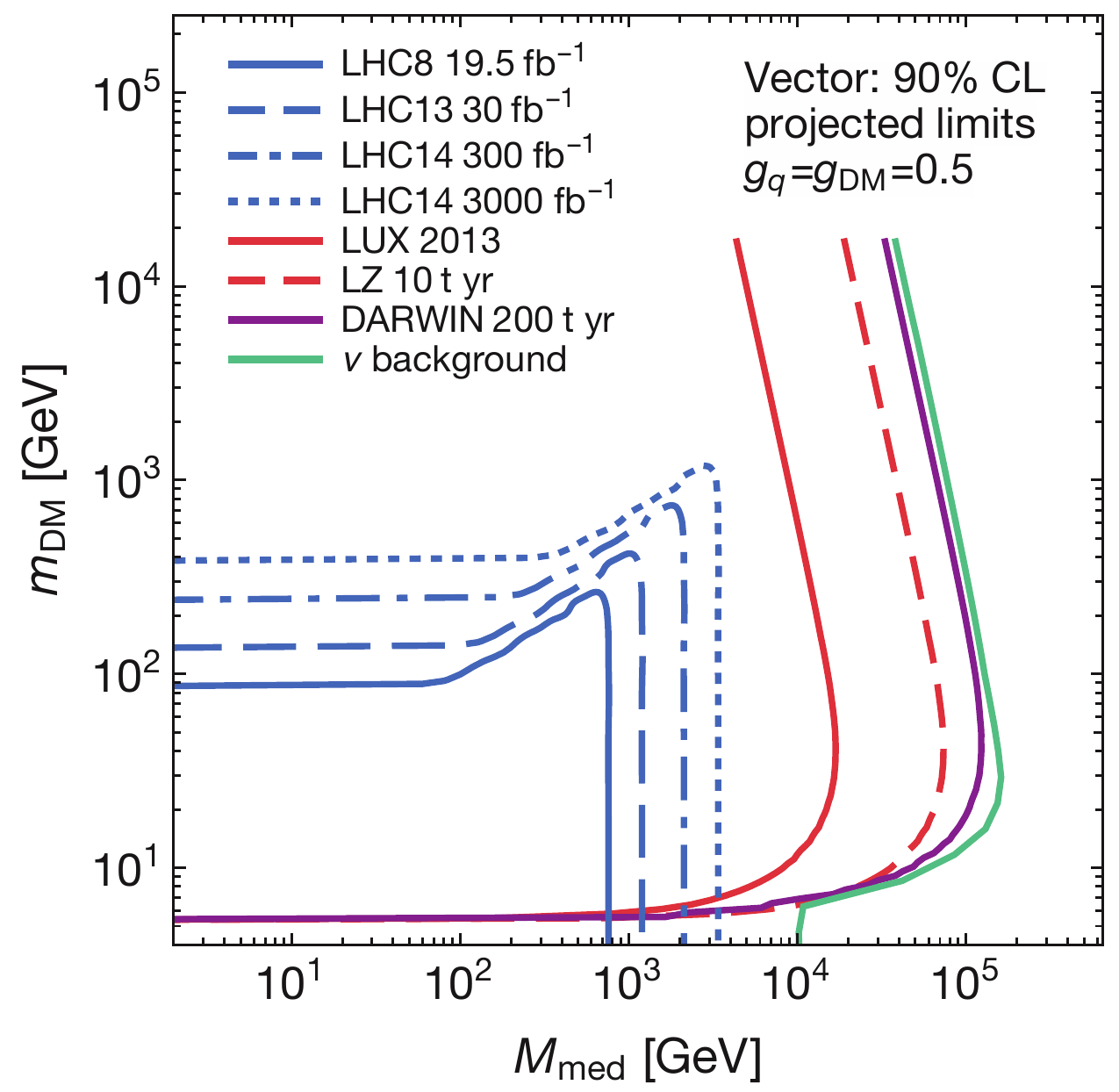}
\includegraphics[width=0.495\columnwidth]{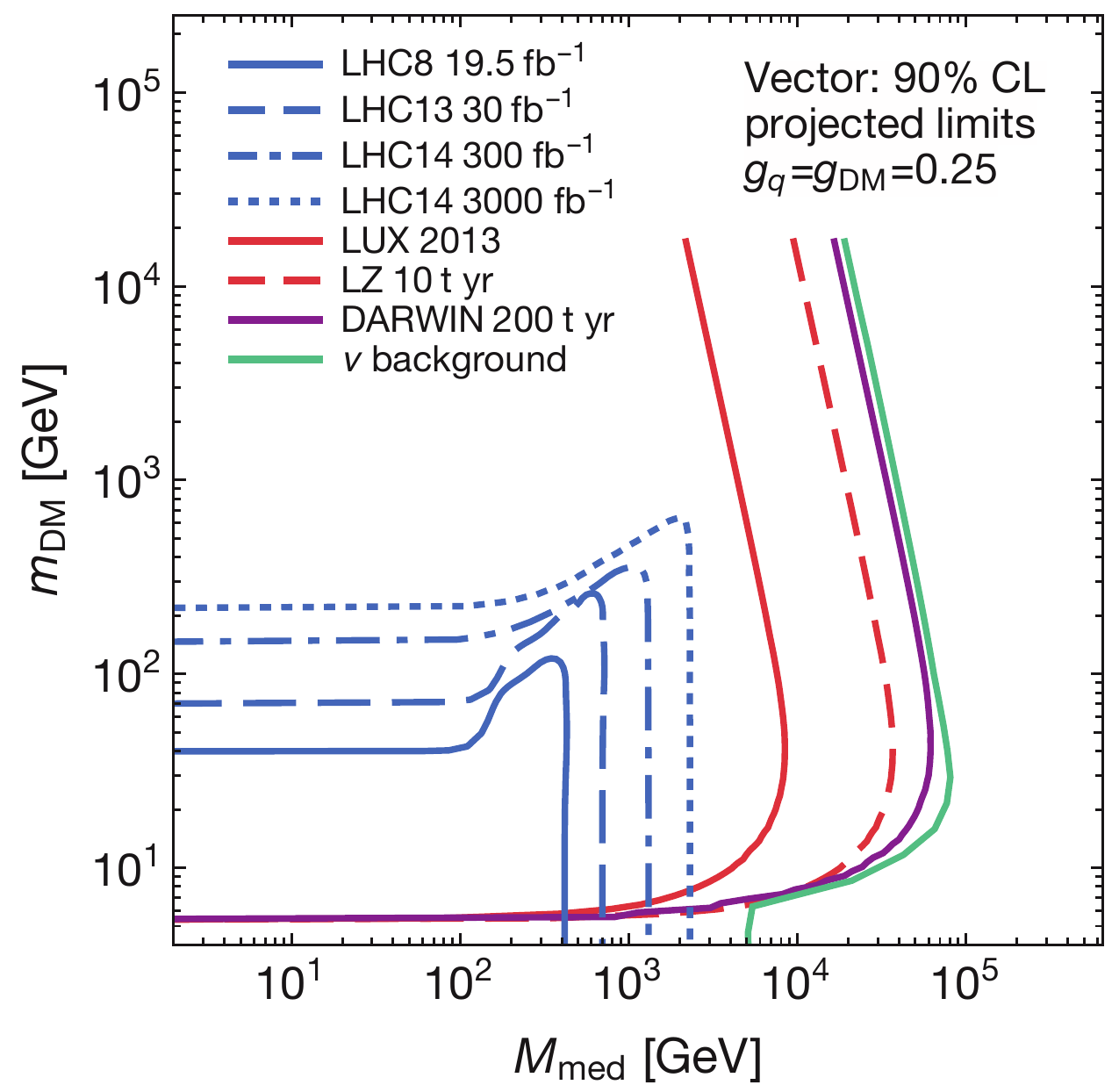}
\caption{Projected limits for the CMS mono-jet search (blue lines) and DD searches by LUX (red line), LZ (red dashed line)
and DARWIN (purple line)
in the $(\mMed, \mDM)$ plane for a vector mediator with the coupling scenarios
$\gq = \gDM= 0.25$, 0.5, 1.0, 1.45.  For reference, the discovery reach of DD experiments accounting for the coherent neutrino scattering background is also displayed (green line). The region to the left of the various curves is
excluded at 90\% CL.}
\label{fig:projV}
\end{figure}

Figures~\ref{fig:projAV} and~\ref{fig:projV} show for the different coupling scenarios the current and projected 
90\%~CL limits for the CMS mono-jet and DD searches in the $(\mMed, \mDM)$ plane for the cases
of an axial-vector mediator and a vector mediator, respectively. The conclusions are the same for the 
projected limits as for the current results. We predict similar complementarity between the collider and 
DD experiments going forward, with LUX, LZ and DARWIN retaining better sensitivity than the mono-jet LHC search for
vector mediators for all but the very low~$\mDM$ region, whereas for the axial-vector mediator the 
mono-jet search extends the reach of the DD experiments. As expected, the overall largest reach in the 
DM parameter space is obtained for the largest coupling scenario $\gq=\gDM=1.45$.
Whereas for vector interactions, shown in Figure~\ref{fig:projV}, none of the projected collider or next generation DD limits
extend beyond the discovery reach, the situation for axial-vector mediators shown in Figure~\ref{fig:projAV} is different. 
For $\gq=\gDM=1.45$, neither collider nor DD searches approach in any region of the $\mMed$-$\mDM$ plane the discovery reach. This changes for the weaker coupling scenarios, where for relatively low $\mDM$ the collider limits approach
the DD discovery reach for $\gq=\gDM=1$, and even go significantly beyond it for $\gq=\gDM\le0.5$.

\begin{figure}[t!]
\centering
\includegraphics[width=0.495\columnwidth]{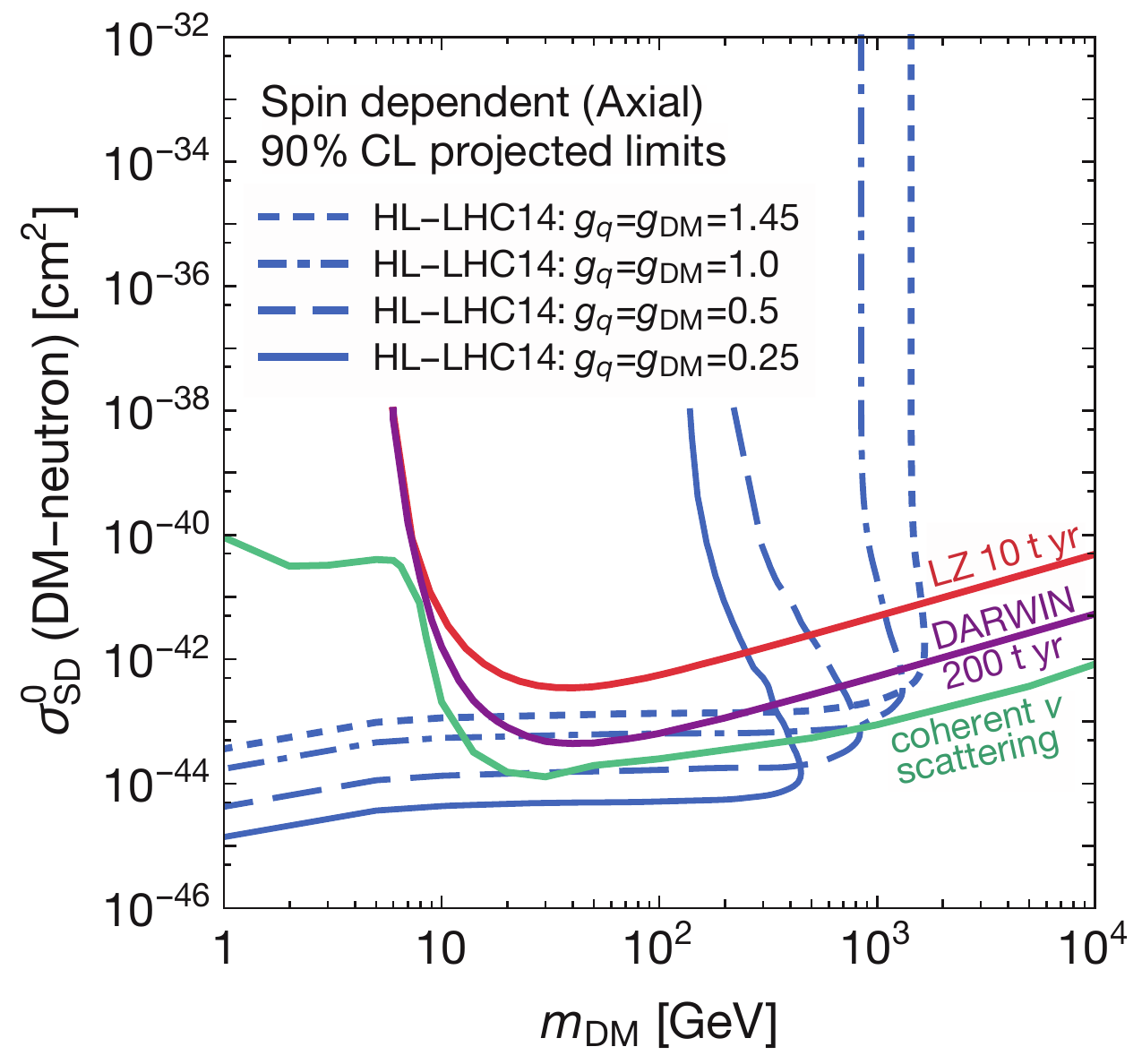} 
\includegraphics[width=0.495\columnwidth]{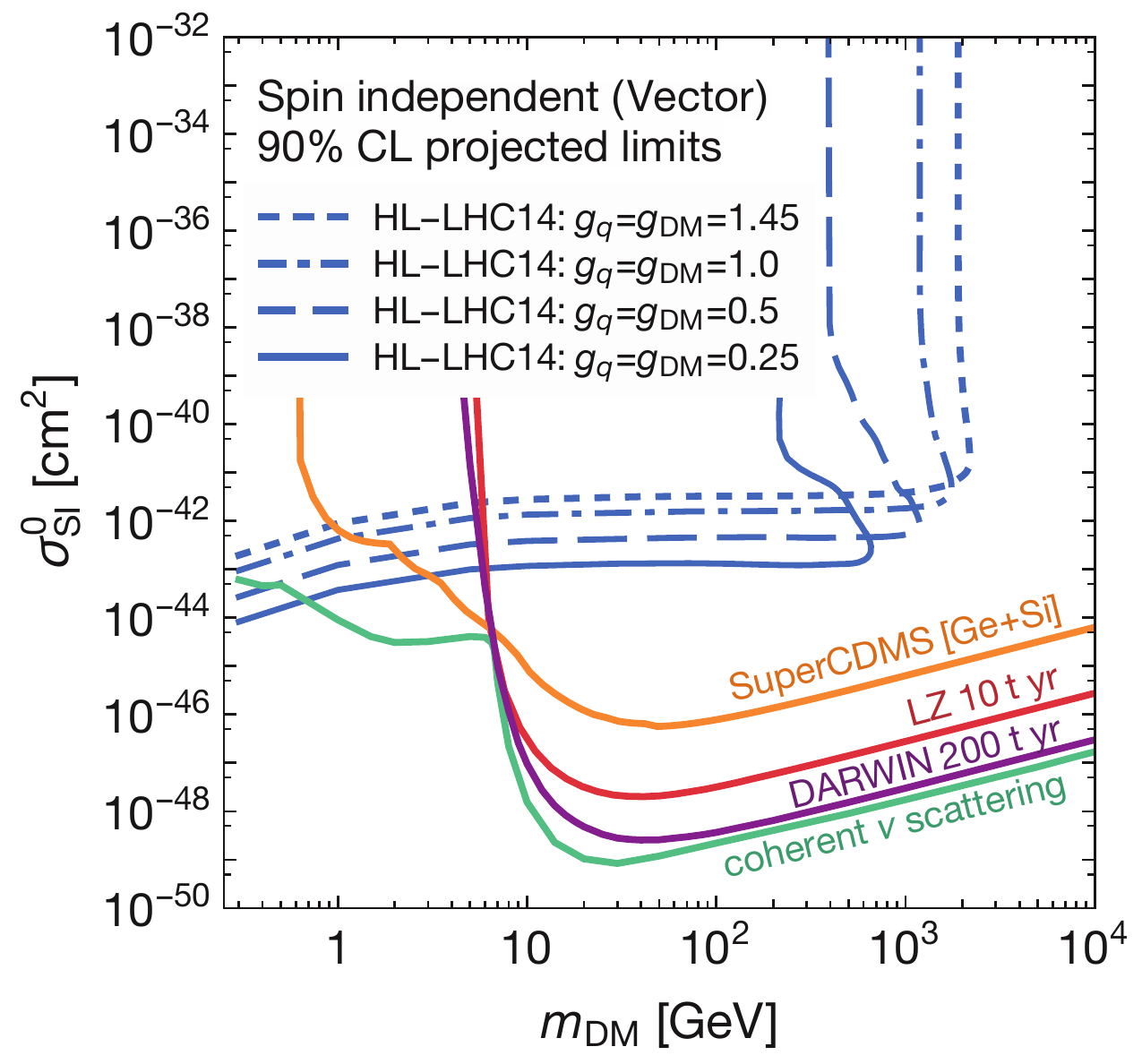}

\caption{Projected 90\% CL limits for the CMS mono-jet search (blue lines), LZ (red lines) and DARWIN (purple lines)
in the cross section vs $\mDM$ plane for SI and SD interactions appropriate for the vector and axial-vector mediators respectively. 
The collider limits are defined for coupling scenarios with $\gq = \gDM = 0.25$, 0.5, 1.0, 1.45. For comparison,
the discovery reach of DD experiments accounting for the neutrino scattering background is also displayed (green lines). 
For the spin-independent interaction we also show a projection of the SuperCDMS limit (orange line).}
\label{fig:proj2}
\end{figure}

The quantitative comparison of the project sensitivities of collider and DD experiments can also be displayed in the traditional
$(\sigma^0_{\rm{SD}}, \mDM)$ and $(\sigma^0_{\rm{SI}}, \mDM)$ planes. 
Figure~\ref{fig:proj2} shows projected limits in these planes for the high-luminosity-LHC (HL-LHC14) scenario of 
3000~fb$^{-1}$ at 14~TeV. Again the four choices of couplings are shown: $\gq=\gDM = 0.25$ and~$1.45$, 
which approximate the extremes of couplings, and the intermediate coupling scenarios of 1.0 and 0.5.

Also shown are the projected limits from LZ and DARWIN assuming a~10 and 200~tonne year exposure respectively,
and the projected spin-independent limits from SuperCDMS assuming a run with 108~Ge and 36~Si 
detectors at SNOLAB~\cite{SuperCDMSproject}. In the case of the spin-independent interactions,
the SuperCDMS projection extends the sensitivity of DD experiments to lower values of $\mDM$,
so its inclusion provides a more complete comparison with the collider limits. 
Similar conclusions regarding the comparison between the MSDM and DD limits can be derived from 
projections in this plane. For spin-independent interactions, the MSDM model with a s-channel vector mediator
adds additional sensitivity only in the very low $\mDM$ region, whereas for spin-dependent interactions
the axial-vector mediator complements the LZ limits very well for DM masses below a few hundred GeV,
and extends sensitivity to the cross section beyond the neutrino limit for DM mass below 10 GeV
in all coupling scenarios. 

Both the choices of planes that compare the projected sensitivities of collider and DD experiments provide
accurate comparisons of the two search strategies in the MSDM on an equal footing. 
Whereas the $(\mMed, \mDM)$ plane might be more familiar to the collider community, 
the $(\sigma^0_{\rm{DD}}, \mDM)$ plane is a more traditional way of displaying this comparison
among the DD community. However, when comparing the two planes care must be taken in the 
interpretation of the relative sensitivities of the different scenarios. For example, whereas in the $(\mMed, \mDM)$
plane the mono-jet limits get stronger with increasing coupling, the same results displayed in the
$(\sigma^0_{\rm{DD}}, \mDM)$ plane show that for DM masses below a few hundred GeV more 
parameter space is ruled out for the weaker coupling scenarios. This is explained by the fact that the 
planes use different observables to benchmark the performance of the search. In one case the 
mediator mass $\mMed$ is the benchmark, whereas in the other case it is the nucleon-WIMP scattering cross section $\sigma^0_{\rm{DD}}$. 
As explained above, the cross section scales as $(\gq\gDM)^2/\mMed^4$  for DD experiments,
and approximately like $(\gq\gDM)^2/(\mMed^4 \Gamma_{\rm{med}})$ for the collider search. 
It is important to take these relations into account when translating between the two planes.  
For the example mentioned above, this implies that, whereas the collider limit on $\mMed$ 
gets stronger with increasing coupling, when taking into account the factor $(\gq\gDM)^2$, 
it rules out less parameter space in $\sigma^0_{\rm{DD}}$ as the coupling increases. Therefore, 
the results displayed in these two planes are fully consistent but represent different ways to benchmark the search. 
Depending on what observable is more relevant for the question at hand, 
either the $(\mMed, \mDM)$ plane or the $(\sigma^0_{\rm{DD}}, \mDM)$ plane might be more appropriate to answer it.  

We emphasize that the results and sensitivity projections presented here are valid for single vector or 
axial-vector mediator exchange, assuming equal coupling to all quarks. Experimentally, DD experiments 
probe a combination of the couplings to $u$ and $d$ quarks for vector exchange and to $u$, $d$ and $s$ 
quarks for axial-vector mediator exchange.
This is in contrast to the mono-jet search. Although the production of the vector or axial-vector mediator is mainly 
sensitive to the coupling to $u$ and $d$ quarks, the mono-jet search is also very dependent on the mediator 
width $\Gamma_{\rm{med}}$, which depends on the couplings to all quarks into which the mediator can decay. 
This therefore motivates one direction in which the MSDM framework should be extended: scenarios with different hypotheses 
for the couplings to various flavours of quarks should be considered, since DD and mono-jet searches probe different weighted combinations of these couplings. 

Other avenues should also be explored to cover a more comprehensive region of DM phenomenology.
These include for instance, scalar and pseudoscalar mediators, t-channel mediators and Majorana fermion or scalar DM scenarios. In addition, the collider searches are also sensitive to the properties of the mediator itself and hence results from 
several different topologies, such as di-jet and multi-jet events with missing transverse energy, can be combined to place limits on the 
MSDM parameter space. This is particularly relevant for scenarios where $\gDM\neq \gq$ (discussed further in Ref.~\cite{Buchmueller:2014yoa}) since one interesting feature of these other channels is that they may probe different combinations of DM and quark couplings. For instance, di-jet searches should be considered as complementary to mono-jet searches since they provide additional constraints on the coupling $\gq$ alone. Other examples are found in Ref.~\cite{Frandsen:2012rk}, where it was demonstrated that orthogonal regions of parameter space can be constrained when mono-jet, mono-photon and di-jet searches are combined. Furthermore, multi-jet plus missing transverse energy topologies, as used to search for supersymmetric particle production at the LHC, will complement and may even improve the sensitivity of the mono-jet search by probing additional final states that are relevant to simplified models that predict significant jet activity in the final sate. Examples are scalar and pseudoscalar models, as discussed in~\cite{Buckley:2014fba,Harris:2014hga,UliEm2015,Buch2015}.  

Additional searches may also allow for MSDM models with more parameters to be constrained. While we have only considered couplings of the mediator to quarks, di-lepton, mono-$Z$, mono-$W$ or invisible Higgs searches could all be employed to constrain the coupling of the mediator to leptons or bosons. This opens the possibility of performing a global fit to a MSDM model, 
incorporating also the constraints from the indirect detection experiments, which are likely to provide important constraints on these MSDM models~\cite{Boehm:2003hm_2}. 
This would be akin to the fits that are performed to specific models of supersymmetry,
and would be particularly useful for characterizing any discovery of a DM signal in the direct
or indirect detection experiments and/or the LHC.

\section{Near-term proposal to compare DM searches based on MSDM models}
\label{sec:proposal}

Based on the discussion presented in Section~\ref{sec:Intro}, we propose the following procedure and 
benchmark plots for the comparison of the collider and DD searches in the study of DM parameter space coverage:

\begin{itemize} 
\item We propose that comparisons be made based on MSDM models as described in Section~\ref{sec:Intro}. 
We initially restrict the proposal to MSDM models where the DM is a Dirac fermion that interacts with a 
vector or axial-vector mediator, with equal-strength couplings to all active quark flavours. 
These models are fully described by four independent parameters.
\item
 We propose to map the collider data into two-dimensional planes, and compare with the results of DD searches
 in both the  ``traditional'' cross section versus $\mDM$ plane (see, e.g., Figures~\ref{fig:XS} and~\ref{fig:proj2}), as well
 as the $(\mMed, \mDM)$ plane (see, e.g., Figures~\ref{fig:projAV} and~\ref{fig:projV}), for the four coupling scenarios
 $\gq = \gDM$ = 0.25, 0.5, 1.0, 1.45. For couplings below $\gq$ =$\gDM$ = 0.25 the present CMS mono-jet search
 does not provide a significant limit, while for $\gq = \gDM$ = 1.45 the width of the mediator becomes larger than its mass. 
 Therefore, the proposed range of coupling scenarios covers the two extreme scenarios ($0.25$ and $1.45$) as well as 
 intermediate cases ($0.5$ and $1.0$).  Depending on the desired application,
 one or even both planes can be used to provide a characterization on equal footing of the absolute 
 and relative performances of collider and DD experiments. 
\end{itemize}
This concrete proposal could be adopted for the near-future data comparisons of collider and DD searches for DM. 
We recommend at the same time to continue the discussion and 
to explore further scenarios and models in order to develop a comprehensive strategy to characterize and
compare these searches in the future and maximise the combined DM particle study potential. While the different collider and DD properties of vector or axial-vector mediators are excellent examples to demonstrate the 
complementarity of the two search strategies, an obvious extension of this proposal would be to also consider 
coupling scenarios where $\gq$ is not universal for all quarks and where $\gDM \neq \gq$, scenarios with scalar and pseudo-scalar mediators as well as t-channel exchanges. For example, a MSDM description with scalar and pseudo-scalar mediators would provide some of the simplest realisations of a non-minimal Higgs sector where the Standard Model Higgs 
interacts and can mix with the (pseudo)-scalar mediators. Therefore, such models provide a direct link with Higgs physics and it might even be possible that there is a common origin of the
electroweak and the DM scales in Nature as it was recently explored in e.g.~\cite{Hambye:2013dgv, Khoze:2014xha}.

\section{Summary}   
\label{sec:summary}

We have focused in this White Paper on a concrete proposal for characterizing
and comparing DM searches in collider and DD experiments, based on the framework of simplified models. 
The results presented here are based on recent work described in~\cite{Buchmueller:2014yoa} and are 
defined in the context of Minimal Simplified Dark Matter (MSDM) models, which have four free parameters:
the mass of the DM particle, $\mDM$, the mass of the mediator, $\mMed$, the coupling of the mediator to the DM, $\gDM$, 
and the coupling of quarks to the mediator, $\gq$. We emphasize that all four parameters are important for translating the 
collider limits into equivalent DD experiment sensitivities. For the example of s-channel vector and axial-vector mediator interactions, 
we show how to characterize the results of searches for DM particles at colliders and direct detection (DD) experiments in
such a way that a comparison between the two approaches can be made on an equal footing. 

Using sensitivity projections from the CMS mono-jet search, LZ, DARWIN and SuperCDMS for future running
scenarios, we compare the limits of these searches in two characteristic planes: those
for $(\mMed, \mDM)$ and $(\sigma^0_{\rm{DD}}, \mDM)$. Both planes provide a straightforward comparison of the two 
search approaches and, depending on the desired application, one or even both planes can be used to provide a 
characterization of the absolute and relative performances of collider and DD experiments.  
This prompts us to formulate a proposal for a better-motivated procedure for comparisons of collider data
with results from direct dark matter search experiments.     

This proposal is based on a particular implementation of simplified models, 
which is only one from several options for developing the comparison of DM searches at collider and DD experiments
beyond the over-simplified EFT interpretation. The extension of the MSDM beyond the assumptions made in this White Paper
will be important to make this approach complete. For instance, coupling scenarios where $\gq$ is not universal for all quarks or where $\gDM \neq \gq$ should be considered, and other mediators should be investigated, such as scalar and pseudoscalar interactions as well as t-channel exchanges. The interpretation framework advocated here represents a potential starting point for going beyond the EFT 
framework, but further additions to the MSDM model, as well as the consideration of alternative approaches, 
will be required to develop a general strategy for comparing collider and DD experiments in the future. 
    
\section*{Note Added}
While finalising this document, we became aware of~\cite{Abdallah:2014hon}, which also addresses aspects of simplified models in order to go beyond  EFT interpretations of DM searches.

\section*{Acknowledgements}

The work of S.M.,~O.B.~and J.E.~is
supported in part by the London Centre for Terauniverse
Studies (LCTS), using funding from the European
Research Council via the Advanced Investigator Grant
267352. The work of J.E.~and M.F.~was supported also in part by the UK STFC
via the research grant ST/J002798/1. The work of G.L.~is partially supported by the DOE Grant~\#DE-SC0010010. The work of C.M.~is supported by the ERC starting grant WIMPs Kairos.

\bibliography{ref}
\bibliographystyle{JHEP}

\end{document}